\documentclass[sigconf, nonacm]{acmart} 
\AtBeginDocument{%
  }

\copyrightyear{2026}
\acmYear{2026}
\acmDOI{}
\acmISBN{}

\usepackage{amsbsy}
\usepackage{siunitx}
\usepackage{colortbl,array,xcolor}
\definecolor{myblue}{rgb}{0.5647,0.6784,0.8824}

\newcommand{\myTableCell}[1]{%
  {\renewcommand{\arraystretch}{1.0}%
  \begin{tabular}[c]{@{}l@{}}#1\end{tabular}}%
}

\begin{document}

\title[picoRing \textit{dual}]{picoRing \textit{dual}: ultralow-power bimanual ring controllers\\for everyday AR}

\author{Hideaki Yamamoto}
\affiliation{%
  \institution{The University of Tokyo}
  \city{Tokyo}
  \country{Japan}
  }
\email{yamamoto-hideaki@akg.t.u-tokyo.ac.jp}
\orcid{0009-0002-8977-8371}

\author{Yifan Li}
\affiliation{%
  \institution{The University of Tokyo}
  \city{Tokyo}
  \country{Japan}
  }
\email{yifan217@akg.t.u-tokyo.ac.jp}
\orcid{0009-0005-9261-6391}

\author{Yoshihiro Kawahara}
\affiliation{%
  \institution{The University of Tokyo}
  \city{Tokyo}
  \country{Japan}
  }
\email{kawahara@akg.t.u-tokyo.ac.jp}
\orcid{0000-0002-0310-2577}

\author{Ryo Takahashi}
\authornote{Correspondence to Ryo Takahashi}
\affiliation{%
  \institution{The University of Tokyo}
  \city{Tokyo}
  \country{Japan}
  }
\email{takahashi@akg.t.u-tokyo.ac.jp}
\orcid{0000-0001-5045-341X}

\renewcommand{\shortauthors}{H. Yamamoto, R. Takahashi, et al.}

\begin{abstract}
Advances in smart eyewear, such as AR glasses and HMDs, bring digital information into outdoor and public situations.
However, traditional handheld controllers for hand interactions remain too bulky and socially awkward for these everyday contexts, failing to match the portability and subtlety required for mobile use.
Ring-based input controllers offer a reliable finger-sensing solution; however, continuous operation with a tiny, battery-constrained ring remains challenging.
Here, we present picoRing \textit{dual}, an ultra-low-power bimanual interface with a pair of ring-wristbands based on inductive coupling.
Unlike the original picoRing \textit{mouse} limited to uni-manual tens of bps slow trackball interaction, picoRing \textit{dual} uses a stable NFC-based ring-to-wristband link to capture fast index-finger motions and thumb-to-index microgestures at approximately \qty{1.4}{kbps}.
By tuning the ring's coil geometry, our system achieves stable NFC against finger bending while complying with NFC's ultra-low power profile.
Therefore, picoRing \textit{dual} with approximately \qty{0.86}{\mW} on the ring side can last approximately $100$ continuous hours on a single charge of the ring's \qty{20}{mAh} battery.
picoRing \textit{dual} has the potential to offer ubiquitous finger input for everyday AR interaction.
\end{abstract}




\begin{teaserfigure}
  \includegraphics[width=\textwidth]{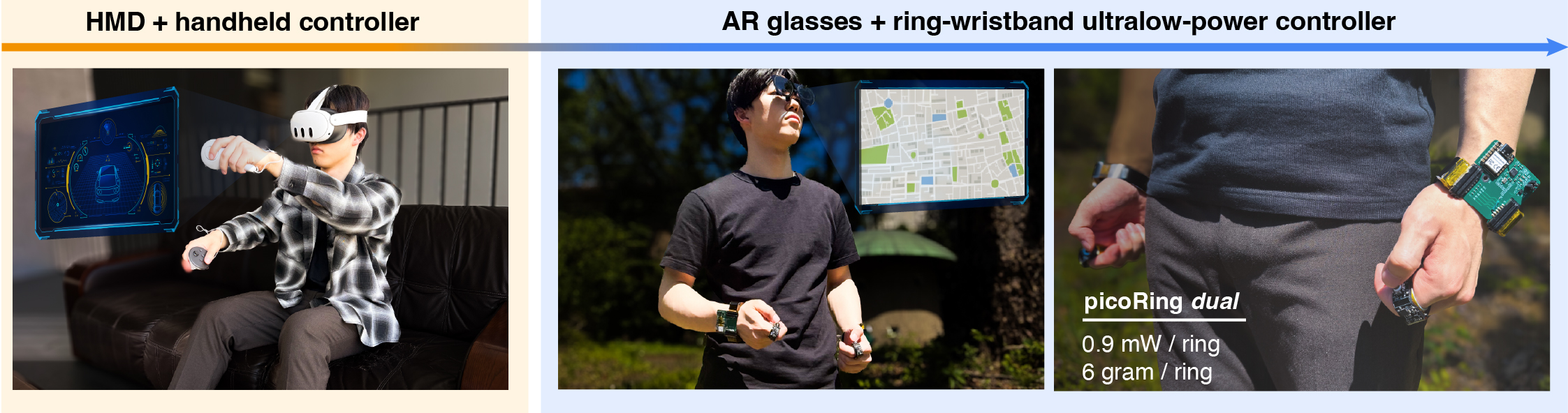}
  \caption{Overview of picoRing \textit{dual}. \textcolor{blue}{\url{https://youtu.be/a0z5SKU4_ug}}}
  \label{fig:overview}
  \Description{A two-part comparison. On the left, labelled HMD plus handheld controller, a person sits indoors wearing a head-mounted display and holding a controller in each hand, with a virtual interface floating beside them. On the right, labelled AR glasses plus ring-wristband ultralow-power controller, a person stands outdoors wearing AR glasses and views a floating navigation map; a close-up shows both hands, each wearing a picoRing dual ring and a wristband reader board, annotated 0.9 milliwatts per ring and 6 grams per ring. An arrow across the top points from the left setup to the right one.}
\end{teaserfigure}


\maketitle

\section{INTRODUCTION}

\begin{table*}[t!]
    \renewcommand{\arraystretch}{1.2}
    \centering
    \caption{Technical comparison of wearable devices for XR interaction. To the best of our knowledge, picoRing \textit{dual} is the first ultralow-power bi-manual dual-ring interface below 1 mW.}
    \begin{tabular}{l|lllll}\toprule
        \textbf{} & 
        \textbf{\myTableCell{Device form factor}} & \textbf{\myTableCell{Power\\consumption\\(mW/device)}} & \textbf{\myTableCell{Wireless\\connection}} & \textbf{\myTableCell{Micro finger\\movement}} & \textbf{\myTableCell{Micro finger\\gesture}}\\\hline\hline
        \multicolumn{6}{l}{\textbf{Dual-handheld, glove, wristband, or fingertip}} \\ \hline
        \rowcolor{myblue!20}Quest Controller~\cite{noauthor_quest_nodate} & Bi-manual by two handhelds & Approx. $27$ & BLE & Pointing & Click and joystick \\
        TapType~\cite{streli_taptype_2022} & Bi-manual by two wristbands & Approx. $10$ & BLE & - & Tap\\ 
        \rowcolor{myblue!20}DigiTouch~\cite{whitmire_digitouch_2017} & Bi-manual by two gloves & - & Gazell & - & Tap and slide \\ 
        BiTipText~\cite{xu_bitiptext_2020} & Bi-manual by two fingertips & - & $\times$ (Wired) & - & Tap \\ \hline
        
        \multicolumn{6}{l}{\textbf{Dual-ring}} \\ \hline
         DualRing~\cite{liang_dualring_2021} & Uni-manual by two rings & - & $\times$ (Wired) & Pointing & Tap and slide \\ 
        \rowcolor{myblue!20}DRG-Keyboard~\cite{liang_drg-keyboard_2023} & Uni-manual by two rings & Approx. $120$ & BLE & - & Tap and slide \\ 
        MouseRing~\cite{shen_mousering_2024} & Uni-manual by two rings & Approx. $10$ & BLE & - & Tap and slide \\ 
        \rowcolor{myblue!20}\textbf{picoRing \textit{dual}} & Bi-manual by two rings & $\mathbf{0.86}$ & Ring-to-wrist NFC & Pointing & Press and scroll \\
        \bottomrule
    \end{tabular}
    \label{tab:comparison}
    \Description{A table comparing eight wearable input devices across five attributes: device form factor, power consumption per device in milliwatts, wireless connection type, micro finger movement support, and micro finger gesture support. The devices are grouped into dual-handheld, glove, wristband, or fingertip devices, and dual-ring devices. picoRing dual, in the last row, has the lowest power consumption at 0.86 milliwatts.}
\end{table*}

Advances in smart glasses, such as head-mounted displays (HMDs) and AR glasses, allow us to use digital information in our daily lives, but how we interact with these devices during our daily lives remains challenging.
Handheld controllers are too bulky for public use, and camera-based hand-tracking systems in AR glasses have issues with continuous use due to \qty{10}{\mW}-class relatively high power consumption and a restricted field of view.
Ring-worn interfaces offer a promising alternative because they can reliably capture nearby finger actions while keeping the palms free~\cite{takahashi_picoring_2024,li_ultra-low-power_2025,wang_computing_2025}.
In particular, dual-ring systems by using both hands can replicate the control of handheld controllers, such as grabbing and stretching 3D objects, or navigating a system menu with one hand while aiming at a target with the other.
However, achieving high-throughput sensing within the tiny, battery-constrained form factor of a ring remains a significant engineering challenge.
While picoRing \textit{mouse}~\cite{takahashi_picoring_2024,li_ultra-low-power_2025} has demonstrated the feasibility of ultra-low-power interaction using inductive coupling, its application is limited to low-bandwidth tasks like cursor navigation. 
The data rate of approximately \qty{50}{bps} is insufficient for capturing the rapid, nuanced finger movements required for smart eyewear input.

To address these challenges, we propose picoRing \textit{dual}, a dual-ring-based bimanual interaction that supports high-throughput finger sensing within an ultra-low-power framework. 
Our system employs an enhanced near-field communication (NFC) between a ring sensor coil and a wristband reader coil on each hand.
By optimizing the coil geometry and utilizing an NFC-based communication scheme, picoRing \textit{dual} achieves a data rate of approximately \qty{1.4}{kbps}—a 28-fold increase over the previous picoRing series~\cite{takahashi_picoring_2024,li_ultra-low-power_2025}, enabling the continuous streaming of fast index-finger motions and thumb-to-index microgestures on both hands simultaneously. 
With this configuration, the ring consuming approximately \qty{860}{\uW} ($\approx \qty{4.2}{\V} \times \qty{205}{\uA}$) can last approximately $100$ hours on a single \qty{20}{mAh} charge.
Therefore, picoRing \textit{dual} provides a privacy-preserving and power-efficient alternative to bulky handheld controllers, paving the way for ubiquitous finger input in everyday AR interaction.
To facilitate replication and provide implementation details, we have open-sourced the current picoRing \textit{dual}: \textcolor{blue}{\url{https://github.com/KawaharaLab/ISWC2026_picoRingdual.git}}.

Our contribution is summarized as follows:
\begin{itemize}
    \item The design of an ultralow-power bimanual dual-ring interface using a ring-to-wristband NFC channel and simple angle optimization of the ring coil. 
    \item The demonstration and technical evaluation of \SI{860}{\uW}-class wireless ring controller named picoRing \textit{dual} toward everyday AR interaction.
\end{itemize}

\section{RELATED WORK}

\subsection{Input techniques for smart eyewear}

As smart eyewear shifts from specific gaming usage to all-day usage, there is a growing demand for subtle and socially acceptable interaction techniques~\cite{dupre_investigating_2025, sapkota_ubiquitous_2021}. 
Traditional input for such devices relies on handheld controllers or optical hand tracking to track mid-air hand poses, as seen in immersive XR.
However, handheld controllers are cumbersome for hands, while optical tracking suffers from occlusion, often failing to capture subtle interactions outside the line of sight~\cite{wolf_performance_2018, novacek_overview_2020}. 
Unlike dynamic mid-air hand poses, subtle inputs allow users to engage in such interactions in public settings without drawing attention or causing arm fatigue. 
To address these limitations, wrist-worn devices such as smartwatches~\cite{lang_multimodal_2023} and smart wristbands~\cite{salemi_parizi_rotowrist_2021} have been investigated for always-available input, employing IMU~\cite{buschek_extending_2018, fan_imu-based_2025}, EMG~\cite{han_dcsnn_2025, cao_finger_2024}, optical~\cite{gong_wristwhirl_2016, salemi_parizi_rotowrist_2021}, pressure~\cite{mcintosh_empress_2016}, and acoustic~\cite{zhuang_reflectrack_2021} sensing. 
Nevertheless, the physical distance between the wrist and the fingers restricts the stable, continuous capture of subtle micro-gestures. 
In contrast, smart rings place sensors directly on the finger, enabling reliable, seamless detection of subtle inputs suitable for everyday use~\cite{takahashi_telemetring_2020,takahashi_picoring_2024,li_ultra-low-power_2025}.

\begin{figure*}[t]
  \centering
  \includegraphics[width=1.0\textwidth]{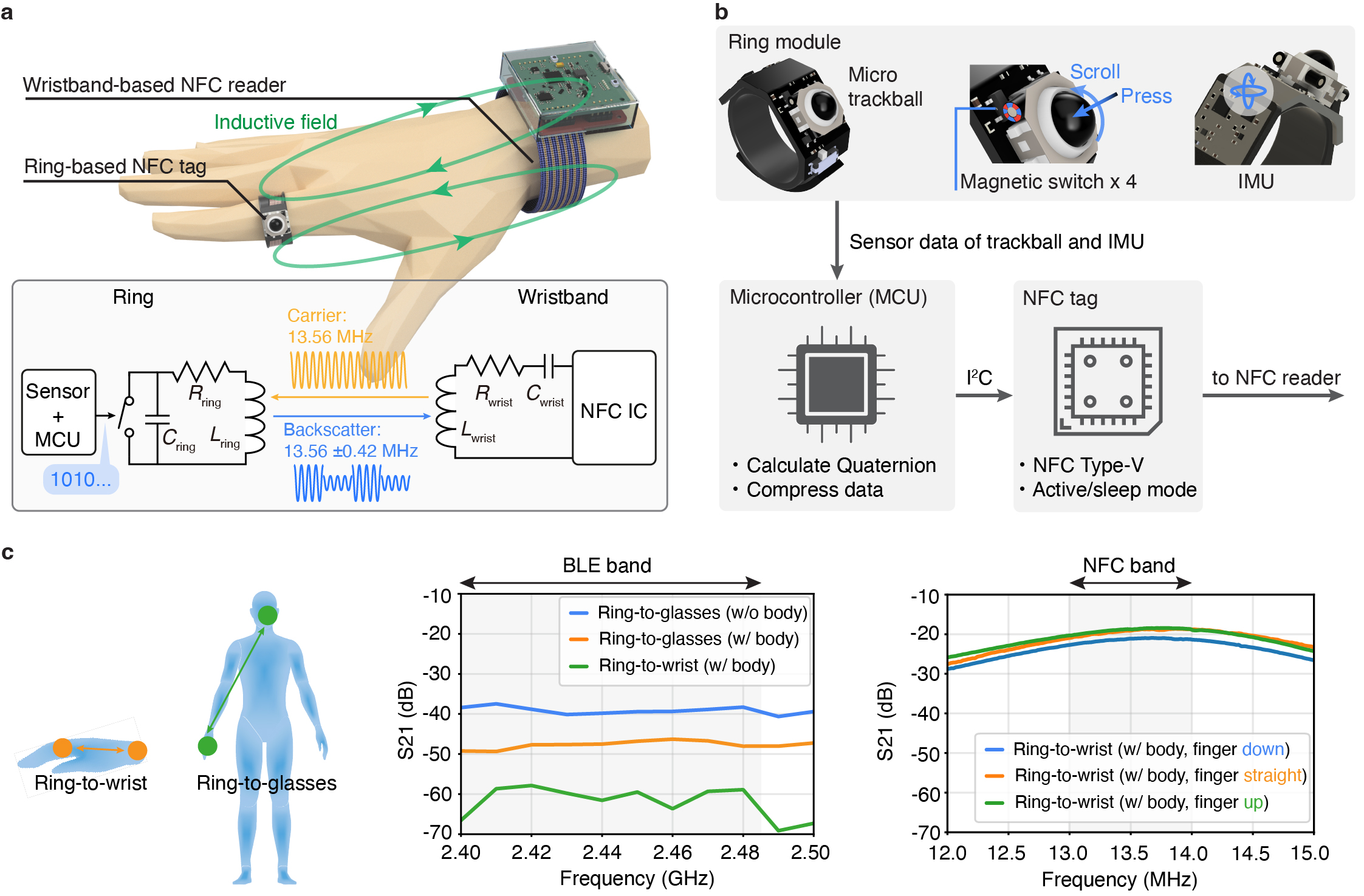}
  \caption{Design overview of picoRing \textit{dual}. (a) Schematic of picoRing \textit{dual}, consisting of a pair of a ring-based NFC tag and a wristband-based NFC reader on the user's both hands. (b) Illustration of the working principle of the ring-based NFC tag to convert the mouse input or motion data into an NFC packet. (c) Communication channel efficiency in the BLE and NFC bands, showing the influence of the human body and varying communication paths (ring-to-wrist and ring-to-glasses).}
  \label{fig:design}
  \Description{Panel (a): a rendering of a hand wearing the ring and the wristband reader, with green loops showing the inductive field linking them; below, a circuit diagram of the ring's resonant tank and the wristband's reader front end, with a 13.56 megahertz carrier travelling to the ring and a backscatter signal at 13.56 plus or minus 0.42 megahertz returning to the wristband. Panel (b): a block diagram of the ring module, showing the micro trackball with scroll and press actions, four magnetic switches, and the IMU, whose data go to the microcontroller, which computes a quaternion and compresses it, then passes it over I2C to an NFC Type-V tag that transmits to the reader. Panel (c): body diagrams of the ring-to-wrist and ring-to-glasses paths, and two line charts of S21 in decibels. In the BLE band chart, ring-to-glasses without the body sits near minus 40 decibels, ring-to-glasses with the body near minus 50, and ring-to-wrist with the body near minus 70. In the NFC band chart, all three ring-to-wrist curves, for finger down, straight, and up, peak near minus 20 decibels around 13.56 megahertz.}
\end{figure*}

\subsection{Dual-ring}
Rings for XR can provide subtle input techniques, including (i) rapid finger pointing~\cite{kim_iris_2024, zhou_one_2023,parizi_auraring_2020,chen_utrack_2013}, and (ii) micro-gestures such as scrolling and clicking~\cite{takahashi_picoring_2024, waghmare_z-ring_2023, sun_thumbtrak_2021,yoon_tring_2016}.
To support these interactions, prior rings utilize motion sensors to enable finger pointing~\cite{shen_mousering_2024, liang_dualring_2021}, in addition to using miniature touchpads~\cite{kubo_ring-type_2022} or trackballs~\cite{li_ultra-low-power_2025} to support microgestures. 
Building on single-ring research, dual-ring usage has emerged to support bimanual interactions. 
By equipping both hands with a ring, users can operate XR environments similarly to traditional handheld VR controllers, eliminating the need for bulky hardware~\cite{gheran_ring_2018}.
While other form factors such as fingertips~\cite{xu_bitiptext_2020}, wristbands~\cite{streli_taptype_2022}, and gloves~\cite{whitmire_digitouch_2017} have been explored for bimanual subtle input, the finger-covered devices often obstruct natural tactile sensations, or wristbands need to rely on external surfaces.
In contrast, dual rings keep the sensitive fingertips entirely unobstructed, ensuring that users can maintain their everyday tactile feedback and seamless hand-to-object interactions while still enabling robust bimanual input in XR.
The dual-ring has two kinds of configurations: two rings on a single hand~\cite{zhou_learning_2022} or one ring on each hand~\cite{gheran_ring_2018}.
For example, two motion rings on the index fingers can detect finger joints such as handwriting~\cite{he_writingring_2025} and typing~\cite{wu_ringbyte_2025}, while placing one ring on each hand mimics the familiar setup of standard dual VR controllers for applications like gaming~\cite{gheran_ring_2018}.
However, achieving the low-power ring controller for everyday XR interaction is still challenging due to power-hungry wireless communication against a ring's tiny battery (see Tab. \ref{tab:comparison}).

\subsection{Near-field communication}

NFC, which is one of the RFID technologies, is designed for short-range identification and payment by wirelessly powering up tags and sending data through load modulation.
Unlike active protocols like BLE that require internal power to constantly broadcast signals, NFC operates through passive inductive coupling and load modulation. 
Instead of emitting its own radio waves, the NFC tag simply alters its impedance to reflect the reader’s magnetic field, allowing the tag side to achieve near-zero power consumption or a completely battery-free form factor.
While early HCI research focused on simplified pairing and tangible interfaces~\cite{hardy_touch_2008,fei_peripheral_2013}, recent work has expanded the NFC’s utility by embedding tags into everyday accessories and toys, ranging from jewelry~\cite{lee_nfcstack_2022,zhao_nfc-wisp_2015}, clothing~\cite{takahashi_meander_2022,takahashi_twin_2021,takahashi_full-body_2025,yamamoto_body-scale_2026}, furniture~\cite{wang_locating_2023,oh_battery-free_2021}, capsule~\cite{zhang_nfcapsule_2022}, patch~\cite{shao_room-temperature_2022,yamagishi_flexible_2024}, and gadgets~\cite{ma_nfinger_2026,li_nfcgest_2025}.
These integrations enable battery-free wireless sensing, specifically for on-demand gesture recognition and physiological monitoring~\cite{zou_nfcrfid-enabled_2025}.
Building on these NFC works, picoRing \textit{dual} introduces an ultralow-power bimanual interface that addresses the battery constraints of ring-sized controllers.

\begin{figure*}[t!]
  \centering
  \includegraphics[width=2\columnwidth]{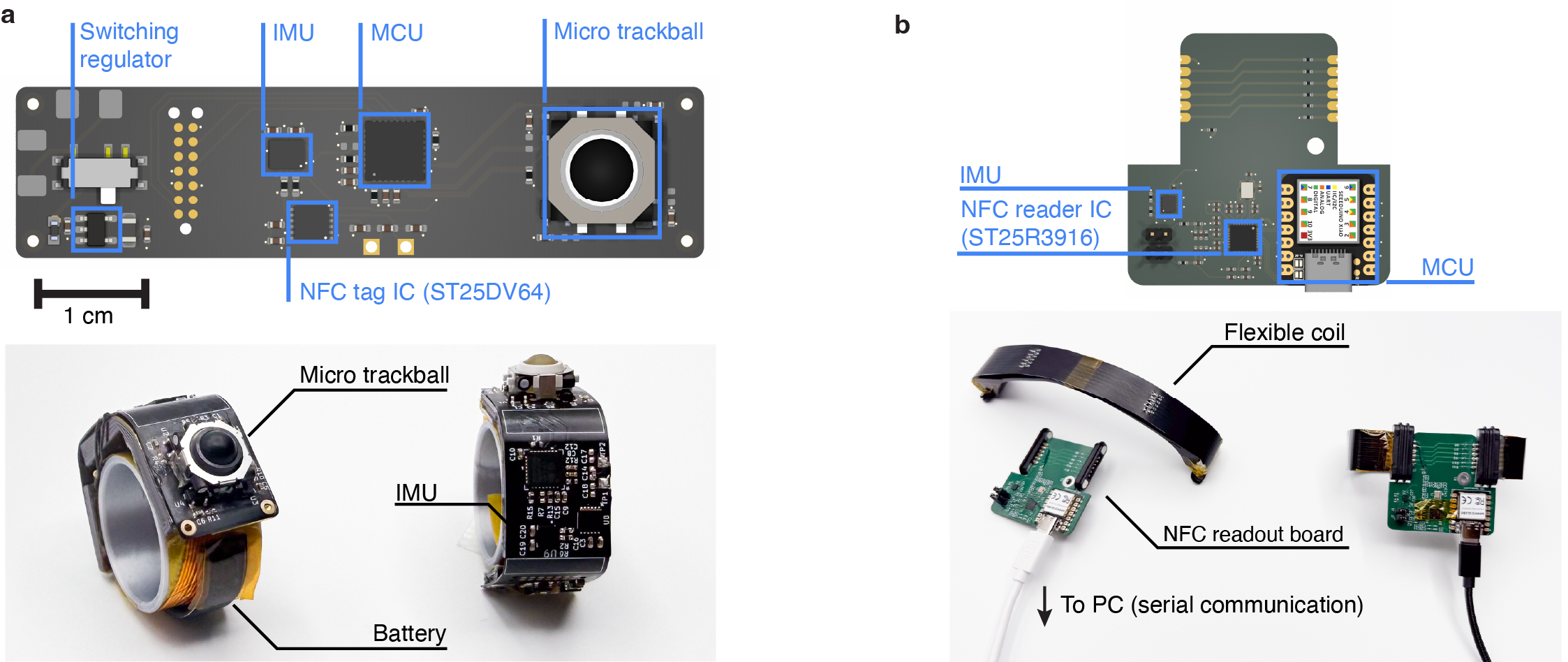}
  \caption{Prototype photograph of our (a) ring-based NFC sensor controller and (b) wristband-based NFC reader.}
  \label{fig:implementation}
  \Description{Two photographs. Panel (a) shows the ring-based NFC sensor controller: a narrow flexible printed circuit board wrapped around a 3D-printed ring base, with a micro trackball, an IMU, a microcontroller, an NFC tag chip, and a coil wound around the ring. Panel (b) shows the wristband-based NFC reader: a flexible coil mounted on a 3D-printed wristband and connected to a readout board carrying an NFC reader chip and a microcontroller.}
\end{figure*}

\section{DESIGN AND IMPLEMENTATION}

picoRing \textit{dual} consists of a ring-based NFC tag module and a wristband-based NFC reader on each hand (see \autoref{fig:design}a). 
The ring module integrates a micro trackball, magnetic switches, an IMU, a low-power microcontroller (MCU), and a custom coil. 
The ring's MCU continuously samples data from the trackball and IMU, calculates quaternions, and compresses the payload before transferring it via I2C to an NFC tag, which transmits the data via passive backscatter through inductive coupling ($13.56 \pm \qty{0.42}{\MHz}$) (see \autoref{fig:design}b). 
By limiting the NFC transmission distance to a nearby wristband, each ring on both hands can utilize the NFC protocol for ultralow-power communication. 
Since the wristband can accommodate a relatively large battery, it acts as a power-sufficient hub that relays the captured interaction data to other devices (e.g., AR glasses) via BLE, allowing the rings to focus entirely on the short-range NFC link.

To confirm the feasibility of the ring-to-wrist inductive link, we evaluated the communication channel efficiency by measuring the $S_{21}$ scattering parameter (the ratio of received to transmitted power) across both the NFC (\qty{13.56}{\MHz}) and BLE (\qty{2.45}{\GHz}) bands. 
Specifically, we connected the chip antennas (EB$\_$NN03-320-M-BT, Ignion) and custom coils to a vector network analyzer (PicoVNA 108) to test two configurations: ring-to-wrist and ring-to-glasses (see \autoref{fig:design}c). 
The results showed that while the BLE ring-to-glasses channel maintains $\qty{-40}{\dB}$ in free space, the nearby dielectric human body causes severe attenuation, dropping the $S_{21}$ to $\qty{-70}{\dB}$. Even at the short ring-to-wrist distance, BLE exhibits low efficiency at around $\qty{-50}{\dB}$ due to body-induced loss. 
In contrast, the short-range NFC ring-to-wrist channel achieves a robust $S_{21}$ of $\qty{-20}{\dB}$, outperforming BLE by over $\qty{30}{\dB}$, though the ring-to-glasses channel becomes noisy below $\qty{-70}{\dB}$ due to far-away distance. 
Furthermore, the NFC link exhibits high tolerance to dynamic hand postures; even when the finger moves up or down, the channel stably sustains an $S_{21}$ between $\qty{-20}{\dB}$ and $\qty{-25}{\dB}$. These results validate that the short-range inductive link is highly effective for ultra-low-power ring operation.
Note that, while the most similar work, picoRing \textit{mouse}~\cite{li_ultra-low-power_2025} is restricted to a slow \qty{50}{bps} datarate due to its frequency-sweeping readout, picoRing \textit{dual} increases the datarate by \num{28} times to approximately \qty{1.4}{kbps} by using a load-modulation-based binary phase shift keying (BPSK) via the single carrier frequency of \qty{13.56}{MHz}.
This enhanced bandwidth enables the rings to continuously stream high-throughput IMU motion data while maintaining only \qty{0.86}{\mW} power consumption on the ring side.

\subsection{Ring}
\label{sec:ring}

\begin{figure*}[t!]
  \centering
  \includegraphics[width=1.0\textwidth]{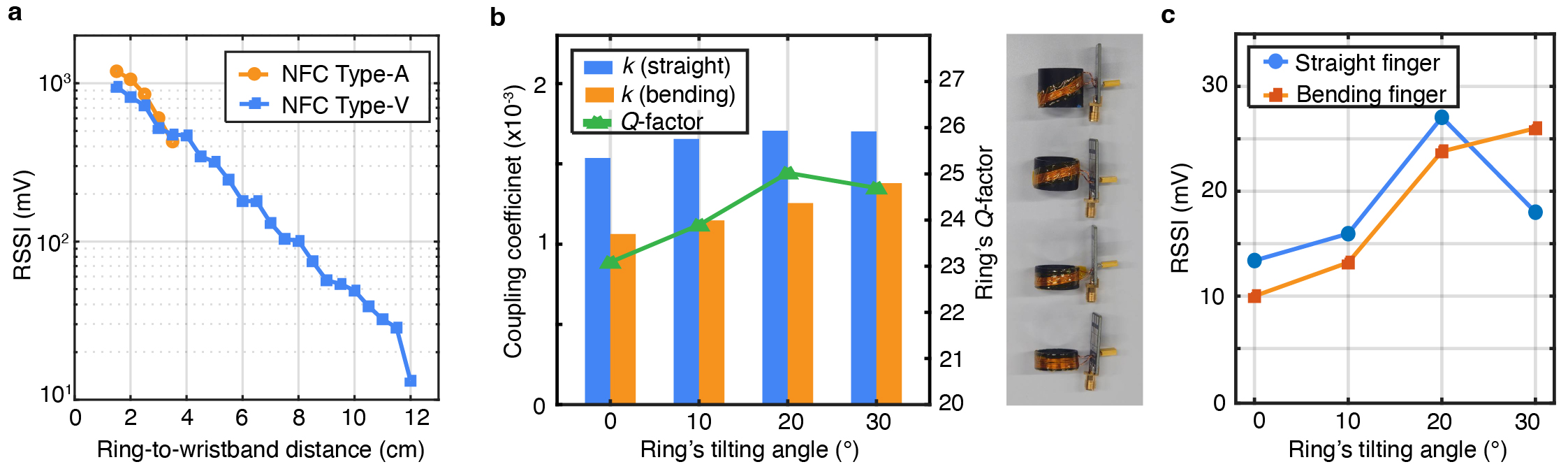}
  \caption{Evaluation of picoRing \textit{dual}. (a) Received Signal Strength Indicator (RSSI) profiles of NFC Type-A and Type-V protocols, demonstrating the extended operating range of Type-V under weak inductive coupling. (b) Measured inductive coupling coefficient ($k$) for straight and bending finger postures alongside the ring's quality factor ($Q$-factor). 
  (c) RSSI performance under straight and bending finger postures across different ring tilting angles, validating that a tilting angle of $20^\circ$ shows the optimal balance of signal strength for natural hand gestures.
  }
  \label{fig:evaluation}
  \Description{Three charts. Chart (a) plots RSSI in millivolts on a logarithmic axis against ring-to-wristband distance from 0 to 12 centimeters, for NFC Type-A and Type-V. The two protocols overlap at short range, but the Type-A trace ends at around 3.5 centimeters, whereas Type-V continues to about 12 centimeters, where it falls to roughly 12 millivolts. Chart (b) is a bar chart of the inductive coupling coefficient k, in units of ten to the minus three, for straight and bending finger postures at ring tilting angles of 0, 10, 20, and 30 degrees, overlaid with a line for the ring's Q-factor on a right-hand axis. Both k values rise with angle, with the bending-finger k rising more steeply, and the Q-factor peaks near 25 at 20 degrees. Photographs of the four tilted coil prototypes are shown alongside. Chart (c) plots RSSI in millivolts against the same tilting angles for straight and bending fingers. Both curves rise to a peak near 20 degrees, where the straight-finger value reaches about 27 millivolts and the bending-finger value about 24 millivolts; at 30 degrees the straight-finger value drops sharply to about 18 millivolts while the bending-finger value continues to rise.}
\end{figure*}

The ring coil consists of an $ 8$-turn resonant coil wounded by $\phi $\qty{0.5}{\mm} copper wire and a flexible PCBs including NFC tag IC (ST25DV64, STM), MCU (STM32U375, STM), IMU (ICM-45605, TDK), switching regulator (LM3670MF, TI), trackball~(EVQWJN007, Panasonic), and 3D-printed ring base (see \autoref{fig:implementation}a).
The ring's MCU samples the raw interaction data from the IMU and micro trackball at \qty{20}{\Hz}, applies a Madgwick filter to compute the finger orientation as a quaternion, and compresses the data payload from float-type $16$~bytes to int-type $8$~bytes ($2$-bytes $\times 4$ quaternion) with $1$-byte trackball state to minimize transmission overhead before transferring it via I2C to the NFC tag IC.
We selected NFC Type-V (ISO/IEC 15693) over standard Type-A or Type-B because Type-V is optimized for mid-range communication. 
Its narrow bandwidth of \qty{13.56 \pm 0.42}{\mega\hertz} increases receiver sensitivity to weak magnetic fields, allowing the system to maintain a stable ring-to-wristband link.

Furthermore, the coil is wound with an inclination of $20^\circ$ relative to the ring's horizontal plane. 
Because most mid-air finger gestures are performed with slightly bent fingers, a parallel coil design would misalign with the wristband reader during interaction. 
Tilting the coil ensures that the loop remains parallel to the wristband when the finger is naturally flexed.
This angle optimization increases the inductive coupling coefficient ($k$) by a factor of $1.2$, as described in \S~\ref{sec:eval_angle}.
As a result, the system maintains a stable link even at a ring-to-wristband distance over \qty{10}{\cm}, which is sufficient for standard hand size~\cite{takahashi_picoring_2024}.
The resonant frequency of the ring coil is adjusted to \qty{13.56}{\MHz} by connecting the chip capacitor of \qty{30}{\pF} in parallel.
In total, $Q$-factor, inductance, power consumption, and datarate of the ring are \qty{25}{}, \qty{2.4}{\uH}, \qty{0.86}{\mW} ($=\qty{4.2}{\V}\times \qty{205}{\uA}$), and \qty{1.4}{kbps} ($=\qty{9}{byte}\times \qty{8}{bit}\times \qty{20}{\Hz}$), respectively.
Note that the reported data rate is an analytical estimate; a detailed empirical evaluation is left for future work.

\subsection{Wristband}
\label{sec:wrist}

The wristband coil consists of a $5$-turned flexible resonant coil mounted on a 3D-printed flexible wristband and the NFC readout board including MCU (XIAO SAMD21, Seeed Studio) and NFC reader IC (ST25R3916, STM) powered by USB PD cable connected to an external PC (see \autoref{fig:implementation}b). 
To recognize the ring's data, the wristband coil captures the load-modulated backscatter signals generated by the ring.
The NFC reader IC demodulates the $\qty{13.56}{\MHz}$ carrier to extract the $\qty{424}{\kHz}$ subcarrier with \qty{26.48}{\kHz} datarate, decoding the BPSK-encoded sensor packet. 
Once decoded, the onboard MCU processes the incoming stream and immediately transmits the compressed orientation and trackball data to the host device via USB cable.
Standard wristband reader coils are typically implemented with a low number of turns and low self-inductance ($L_{\text{wrist}}$) to prioritize matched impedance over standard short ranges. 
In contrast, we need to increase the number of turns to maximize $L_{\text{wrist}}$, thereby boosting the reader's sensitivity to subtle load changes caused by the ring.
When the ring tag modulates its impedance ($\Delta Z_{\text{ring}}$), the input impedance looking into the wristband coil ($Z_{\text{in}}$) changes according to the mutual inductance ($M = k\sqrt{L_{\text{ring}}L_{\text{wrist}}}$), expressed as: $Z_{\text{in}} = Z_{\text{wrist}} + (\omega M)^2/Z_{\text{ring}}$
where $Z_{\text{wrist}}$ and $Z_{\text{ring}}$ are the inherent impedance of the wristband and ring coil, respectively, and $\omega$ is the angular frequency of the carrier signal.
By maximizing $L_{\text{wrist}}$, the $(\omega M)^2$ scales significantly higher. 
This amplification ensures that even when $k$ is low due to distance or misalignment, the variation in $Z_{\text{in}}$ remains large enough to stably detect and demodulate.
In total, $Q$-factor, inductance, and power consumption of the wristband are \qty{9}{}, \qty{3.2}{\uH}, and \qty{170}{\mW} ($=\qty{3.3}{\V}\times \qty{51}{\mA}$), respectively.
The power consumption could be reduced below \qty{50}{\mW} by using a low-power NFC reader IC (ST25R200, ST).

\section{TECHNICAL EVALUATION}

\begin{table}[t!]
    \centering
    \caption{Power consumption of the ring coil hardware.}
    \begin{tabular}{lll}\toprule
        \textbf{Component} & \textbf{\myTableCell{NFC (\si{\uW})}}  & \textbf{\myTableCell{BLE (\si{\uW})}}\\\hline
        MCU (STM32U375) & $169$ & $169$ \\
        Trackball module (4 $\times$ CT8132) & $7.9$ & $7.9$ \\
        IMU (ICM-45605) & $396$ & $396$ \\
        Power management (LM3670MF) & $63$ & $63$ \\
        \textbf{Wireless communication} & \textbf{\myTableCell{$\mathbf{130}$\\(ST25DV64)}} & \textbf{\myTableCell{$\mathbf{11,520}$\\(nRF52840)}} \\ \hline
        \textbf{Estimated total} & $>765$ & $>12,200$\\
        \rowcolor{myblue!20}\textbf{Measured total@4.2~V} & $\mathbf{860}$ & $\mathbf{28,000}$\\\bottomrule
    \end{tabular}
    \label{tab:power}
    \Description{A table listing the measured power consumption of each ring component in microwatts under NFC and BLE. The rows are the microcontroller, the trackball module, the IMU, the power management chip, and the wireless communication chip. Wireless communication dominates the difference: 130 microwatts for the NFC tag versus 11,520 microwatts for the BLE radio. The measured totals at 4.2 volts are 860 microwatts for NFC and 28,000 microwatts for BLE.}
\end{table}

\begin{figure*}[ht!]
  \centering
  \includegraphics[width=1.0\textwidth]{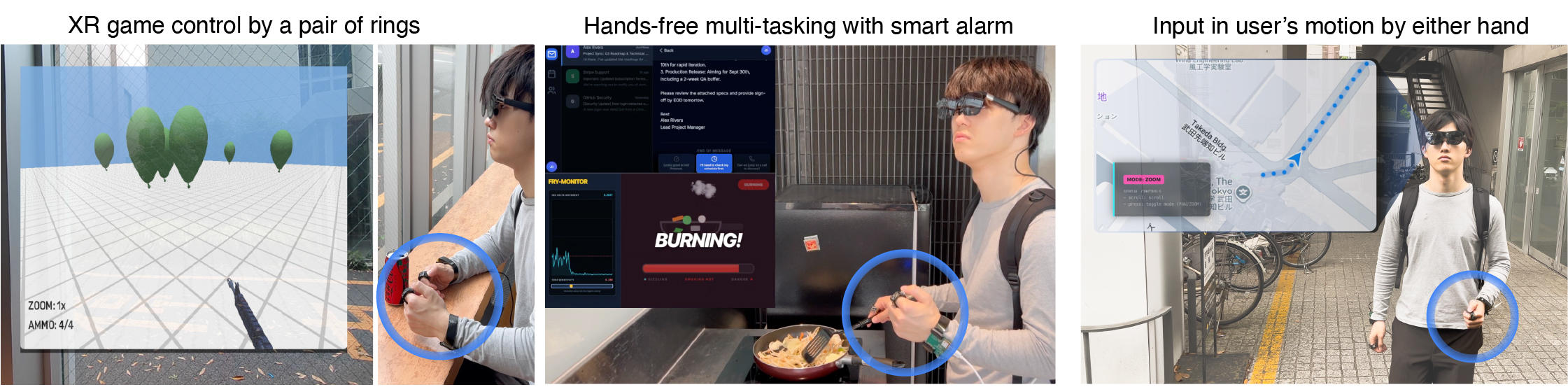}
  \caption{Application examples of picoRing \textit{dual}.}
  \label{fig:app}
  \Description{Three photographed scenarios, each pairing the AR view with the user's hand, circled in blue, wearing the ring and wristband. Left, labelled XR game control by a pair of rings: a first-person shooting scene with balloon targets and a zoom and ammunition readout, played by a seated user. Middle, labelled hands-free multi-tasking with smart alarm: an e-mail thread shown alongside a frying monitor that warns BURNING while the user cooks at a kitchen. Right, labelled input in user's motion by either hand: a navigation map that the user pans and zooms while walking outdoors with the hand at their side.}
\end{figure*}

\subsection{Operation Time using BLE vs NFC}

Based on the measured power consumption of the ring hardware (\autoref{tab:power}), we evaluated the operational ring lifetime when powered by a miniature \qty{20}{mAh} lithium-ion curved battery designed for smartring. 
When configuring the ring for continuous active tracking, the measured power consumption at \qty{4.2}{V} significantly differs between the two wireless protocols (BLE and NFC). 
Operating the device entirely over a standard BLE active connection with the transmission power of \qty{0}{dBm} consumes approximately $\qty{28,000}{\micro\W}$ ($\qty{28}{\mW}$), draining the \qty{20}{mAh} battery in just \qty{4}{\hour}.
Even though we decrease the input power to \qty{-40}{dBm}, the power consumption remains almost the same as that of \qty{0}{dBm}.
In contrast, NFC drastically decreases the power below \qty{900}{\uW}, enabling the continuous operation over $100$ hours.
Therefore, picoRing \textit{dual} allows the ring to achieve an operational lifespan comparable to conventional handheld controllers, supporting 4 to 5 days of continuous use on a single charge. 
Furthermore, with typical daily use of 3 to 4 hours, the battery life extends to approximately one month, making it highly practical for long-term everyday use.

\subsection{Tilted Ring Optimization for NFC Type-V}
\label{sec:eval_angle}

To maximize communication range between ring and wristband, we selected NFC Type-V over standard Type-A after evaluating their Received Signal Strength Indicator (RSSI) profiles across distances from \qty{1.5}{\cm} to \qty{12.0}{\cm} (\autoref{fig:evaluation}a). 
While the Type-A protocol loses connection beyond \qty{3.5}{\cm} due to its high RSSI demodulation threshold around \qty{400}{\mV_{rms}}, Type-V successfully maintained a stable wireless link up to \qty{12.0}{\cm}, operating reliably down to a minimal RSSI threshold of only around \qty{10}{\mV_{rms}}.
This $3.4$-fold increase in communication distance is driven by Type-V's narrow signal bandwidth, which significantly lowers the receiver's noise floor and threshold sensitivity. 
As a result, Type-V can successfully decode attenuated signals, making it ideal for the restricted, low-coupling finger-to-wrist path.

To further stabilize the link during dynamic hand movements, we optimize the ring's coil geometry by evaluating four prototypes with tilting angles ($\phi$) from $0^\circ$ to $30^\circ$ under straight and bent finger postures (see \autoref{fig:evaluation}bc). 
Increasing the angle generally increases the coil's area, increasing the ring's $L$ or $Q$-factor besides the $k$ (see \autoref{fig:evaluation}b). 
We confirm that $\phi = 20^\circ$ is the best among the four tilting angles because $\phi = 20^\circ$ shows high RSSI over \qty{20}{\mV_{rms}} for the straight/bending posture.

\section{APPLICATION EXAMPLES}

picoRing \textit{dual} offers a significant advantage in providing long-term continuous operation on a single charge, while capturing subtle thumb-to-index microgesture and index finger movements. 
Therefore, its main application is ubiquitous finger input for smart eyewear, particularly when integrated with augmented reality (AR) glasses. 
\autoref{fig:app} shows some application examples of picoRing \textit{dual}.
First, in AR gaming, picoRing \textit{dual} enables reliable, occlusion-free bimanual control—such as combining one hand's trackball for locomotion with the other's IMU for aiming—offering a handheld-controller experience in a wearable form factor.
Next, picoRing \textit{dual} enables hands-free multitasking during everyday activities like cooking. 
While the user's hands are physically occupied such as holding a spatula, the rings allow them to manage background digital tasks (e.g., checking e-mail) without disrupting their workflow.
When the ring's IMU detects no hand movement for an extended period while the user is distracted by other tasks, the ring can provide alerts to refocus the user's attention.
Lastly, picoRing \textit{dual} is available for outdoor mobile scenarios, such as interacting with an AR navigation map while walking.
Traditional camera-based hand tracking often fails outdoors due to dynamic ambient lighting, and holding a smartphone while walking introduces safety and ergonomic concerns. 
In contrast, \textit{picoRing dual} enables private, subtle interactions directly inside a pocket or at the user's side.

\section{CONCLUSION AND DISCUSSION}

This paper presented picoRing \textit{dual}, a dual-ring-based ultralow-power bimanual interaction designed to overcome the severe power and form-factor constraints of finger-worn devices. 
By constructing a short-range data communication between the ring and a nearby wristband through the NFC-based inductive link, we shifted the heavy power burden of long-range active radios away from the smartring. 
Driven by an optimized high-$L$ coil and a low-overhead data compression scheme, the ring module achieves a continuous tracking frame rate of \qty{20}{fps} at an ultra-low power consumption of just \qty{0.86}{\mW}. 
Powered by a miniature \qty{20}{mAh} battery, picoRing \textit{dual} extends the ring’s operational runtime to \qty{100}{\hour} ($4$ days), enabling a $25$-fold increase compared to standard BLE setup.

While the current prototype focuses only on the optimization of ring-to-wristband links, the total AR interaction system still has some limitations.
First, user studies about the input accuracy and wearing comfort will improve the wearability and practicality of picoRing \textit{dual} design across various daily situations.
Second, the communication link between the wristband and eyewear is still poor due to the dielectric body if we use standard antenna approaches.
By using metamaterial-based~\cite{tian_wireless_2019} or meander-based~\cite{takahashi_meander_2022} near-field power and data networking, energy-efficient, stable signal propagation from the wristband to glasses would be enabled. 
Finally, the combination of smart eyewear and ring interaction could also benefit other domains such as robotics manipulation and daily healthcare.

\begin{acks}
This work was supported by JST JPMJCS25N4, JPMJAP2401, JSPS 26K02956, 26K23833, 22K21343, and Asahi Glass Foundation. 
\end{acks}

\bibliographystyle{ACM-Reference-Format}
\bibliography{references}

\appendix
\end{document}